\documentclass[showpacs,amsmath,amssymb]{revtex4}
\usepackage{indentfirst}
\usepackage{amssymb}
\usepackage{multirow}
\usepackage{graphicx,color,dcolumn,booktabs,bm}
\begin{document}
\title{\bf Bottom ${\bf (70,1^-)}$ baryon multiplet}
\author{S.M. Gerasyuta}
\email{gerasyuta@SG6488.spb.edu}
\author{E.E. Matskevich}
\email{matskev@hep.phys.spbu.ru}
\affiliation{Department of Physics, St. Petersburg State Forest Technical
University, Institutski Per. 5, St. Petersburg 194021, Russia}
\begin{abstract}
The masses of negative parity $(70,1^-)$ bottom nonstrange baryons
are calculated in the relativistic quark model. The relativistic
three-quark equations of the $(70,1^-)$ bottom baryon multiplet
are derived in the framework of the dispersion relation technique.
The approximate solutions of these equations using the method based
on the extraction of leading singularities of the amplitude are
obtained. The masses of 21 baryons are predicted.
\end{abstract}
\maketitle
\section{Introduction.}
Hadron spectroscopy has always played an important role in the revealing
mechanisms underlying the dynamic of strong interactions.
At low energies, typical for baryon spectroscopy, QCD does not admit
a perturbative expansion in the strong coupling constant. In 1974
't Hooft \cite{1} suggested a perturbative expansion of QCD in terms of the
parameter $1/N_c$ where $N_c$ is the number of colors. This suggestion
together with the power counting rules of Witten \cite{2} has lead to the
$1/N_c$ expansion method which allows to systematically analyse baryon
properties.

In the series of papers \cite{3, 4, 5, 6, 7} a practical treatment of relativistic
three-hadron systems have been developed. The physics of three-hadron
system is usefully described in term of the pairwise interactions
among the three particles. The theory is based on the two principles
of unitarity and analyticity, as applied to the two-body subenergy
channels. The linear integral equations in a single variable are obtained
for the isobar amplitudes. The coupled integral equations are solved in
terms of simple algebra.

In our papers \cite{8, 9} relativistic generalization of the three-body
Faddeev equations was obtained in the form of dispersion relations in the
pair energy of two interacting particles. The mass spectrum of $S$-wave
baryons including $u$, $d$, $s$-quarks was calculated by a method based
on isolating the leading singularities in the amplitude. We searched for
the approximate solution of integral three-quark equations by taking
into account two-particle and triangle singularities, all the weaker ones
being neglected. If we considered such an approximation, which corresponds
to taking into account two-body and triangle singularities, and defined
all the smooth functions of the middle point of the physical
region of Dalitz-plot, then the problem was reduced to the one of solving
a system of simple algebraic equations.

In our paper \cite{10} the construction of the orbital-flavor-spin wave functions
for the $(70,1^-)$ multiplet are given. We deal with a three-quark system
having one unit of orbital excitation. The orbital part of wave function
must have a mixed symmetry. The spin-flavor part of wave function must
have the same symmetry is order to obtain a totally symmetric state in the
orbital-flavor-spin space. The integral equations using the
orbital-flavor-spin wave functions was constructed. It allows to calculate
the mass spectra for all baryons of $(70,1^-)$ multiplet. The $15$
resonances are in good agreement with experimental data. We have
predicted $15$ masses of baryons. In our model the four parameter
are used: gluon coupling constants $g_{+}$ and $g_{-}$ for the various
parity, cutoff energy parameters $\lambda$, $\lambda_s$ for the
nonstrange and strange diquarks.

In the paper \cite{11} the relativistic three-quark equations of the
excited $(70,1^-)$ charmed baryons are found in the framework of
the dispersion relation technique.
The $(70,1^-)$ charmed baryon multiplet has $23$ baryons with different
masses. The $6$ resonances are in good agreement with the experimental data.
We have predicted $17$ masses of charmed excited baryons.

In the framework of the proposed approximate method of solving the
relativistic three-particle problem, we have obtained a
spectrum of $P$-wave bottom baryons. This paper is a generalization of
our works \cite{10, 11}.

The paper is organized as follows. After this introduction, we consider
the derivation of relativistic generalization of the Faddeev equations
for the example of $S$-wave bottom state $\Omega_{bbb}$ with $J^P=\frac{3}{2}^+$.

In Sect. III we discuss
the construction of the orbital-flavor-spin wave functions for the
$(70,1^-)$ bottom multiplets.

In Sect. IV the relativistic three-quark equations are obtained in
the form of the dispersion relation over the two-body subenergy.

In Sect. V the systems of equations for the reduced amplitudes are
derived.

Section VI is devoted to the calculation results for the mass spectrum
of the $(70,1^-)$ bottom multiplet (Tables I-IV).

\section{Brief introduction of relativistic Faddeev equations.}

We consider the derivation of the relativistic generalization of the
Faddeev equation for the example of the $S$-wave bottom state $\Omega_{bbb}$
with $J^P=\frac{3}{2}^{+}$. This is the simplest
example because this state consists of the three similar quarks.

We take into account only pair quark interactions. The state has the three
similar channels consisting of the diquark with the quantum number $J^P=1^+$
(in the color state $\bar 3_c$) and the third quark (in the color state $3_c$).
The $3Q$ baryon state $\Omega_{bbb}$ is constructed as color singlet. Suppose
that there is a $\Omega_{bbb}$ current which produces three $b$ quarks (Fig. 1a).
Accounting all possible pair interactions lead to the diagrams, which
can be grouped in accordance with the latest interacting pair of
particles (Fig. 1b-1d). The total amplitude can be represented as a sum of
diagrams. Taking into account the equality of all pair interactions of
bottom quarks in the state with $J^P=1^+$, we obtain the corresponding
equation for the amplitudes:

\begin{equation}
\label{1}
A_1 (s, s_{12}, s_{13}, s_{23})=\lambda+A_1 (s, s_{12})+
A_1 (s, s_{13})+A_1 (s, s_{23})\, . \end{equation}

\noindent
Here the $s_{ik}$ are the pair energies squared of particles $i$ and $k$, and $s$
is the total energy squared of the system. Using the diagrams of Fig. 1, it is
easy to write down a graphical equation for the function $A_1 (s, s_{12})$
(Fig. 2). For this we need in an explicit form of the amplitude of the pair
quark interaction. We write this amplitude for the diquark with $J^P=1^+$ in the form:

\begin{equation}
\label{2}
a_1(s_{12})=\frac{G^2_1(s_{12})}
{1-B_1(s_{12})} \, ,\end{equation}

\begin{equation}
\label{3}
B_1(s_{12})=\int\limits_{4m^2}^{\Lambda(12)}
\, \frac{ds'_{12}}{\pi}\frac{\rho_1(s'_{12})G^2_1(s'_{12})}
{s'_{12}-s_{12}} \, ,\end{equation}

\begin{eqnarray}
\label{4}
\rho_1 (s_{12})&=&
\left(\frac{1}{3}\, \frac{s_{12}}{4m^2}+\frac{1}{6}\right)
\left(\frac{s_{12}-4m^2}{s_{12}}\right)^{\frac{1}{2}} \, .
\end{eqnarray}

\noindent
Here $G_1(s_{12})$ is the vertex function of a diquark with $J^P=1^+$.
$B_1(s_{12})$ is the Chew-Mandelstam function \cite{12} and $\rho_1 (s_{12})$
is the phase space for a diquark with $J^P=1^+$.

The pair quarks amplitudes $QQ \to QQ$ are calculated in the framework of
the dispersion $N/D$ method with the input four-fermion interaction
\cite{13, 14} with the quantum numbers of the gluon.

The four-quark interaction is considered as an input:

\begin{eqnarray}
\label{5}
 & g_V \left(\bar q \lambda I_f \gamma_{\mu} q \right)^2 +
2\, g^{(s)}_V \left(\bar q \lambda I_f \gamma_{\mu} q \right)
\left(\bar s \lambda \gamma_{\mu} s \right)+
g^{(ss)}_V \left(\bar s \lambda \gamma_{\mu} s \right)^2+\nonumber\\
 & 2\, g^{(c)}_V \left(\bar q \lambda I_f \gamma_{\mu} q \right)
\left(\bar c \lambda \gamma_{\mu} c \right)+
2\, g^{(sc)}_V \left(\bar s \lambda \gamma_{\mu} s \right)
\left(\bar c \lambda \gamma_{\mu} c \right)+\nonumber\\
 & g^{(cc)}_V \left(\bar c \lambda \gamma_{\mu} c \right)^2+
2\, g^{(b)}_V \left(\bar q \lambda I_f \gamma_{\mu} q \right)
\left(\bar b \lambda \gamma_{\mu} b \right)+\nonumber\\
 & 2\, g^{(sb)}_V \left(\bar s \lambda \gamma_{\mu} s \right)
\left(\bar b \lambda \gamma_{\mu} b \right)+\nonumber\\
 & 2\, g^{(cb)}_V \left(\bar c \lambda \gamma_{\mu} c \right)
\left(\bar b \lambda \gamma_{\mu} b \right)+
g^{(bb)}_V \left(\bar b \lambda \gamma_{\mu} b \right)^2
 \, . & \end{eqnarray}

\noindent
Here $I_f$ is the unity matrix in the flavour space $(u, d)$. $\lambda$ are
the color Gell-Mann matrices. Dimensional constants of the four-fermion
interaction $g_V$, $g^{(s)}_V$, $g^{(ss)}_V$, $g^{(c)}_V$, $g^{(sc)}_V$,
$g^{(cc)}_V$, $g^{(b)}_V$, $g^{(sb)}_V$, $g^{(cb)}_V$, and $g^{(bb)}_V$
are parameters of the model. At $g_V =g^{(s)}_V =g^{(ss)}_V=g^{(c)}_V
=g^{(sc)}_V=g^{(cc)}_V=g^{(b)}_V=g^{(sb)}_V=g^{(cb)}_V=g^{(bb)}_V$
the flavour $SU(3)_f$ symmetry occurs.
The $s$, $c$, and $b$ quarks violate the flavour $SU(3)_f$ symmetry. In order to
avoid additional violation parameters we introduce the scale of the
dimensional parameters \cite{14}:

\begin{eqnarray}
\label{6}
 & g=\frac{m^2}{\pi^2}g_V=\frac{(m+m_s)^2}{4\pi^2}g_V^{(s)}=
\frac{m_s^2}{\pi^2}g_V^{(ss)}=\frac{(m+m_c)^2}{4\pi^2}g_V^{(c)}=
\nonumber\\
 & \frac{(m_s+m_c)^2}{4\pi^2}g_V^{(sc)}=
\frac{m_c^2}{\pi^2}g_V^{(cc)}=\frac{(m+m_b)^2}{4\pi^2}g_V^{(b)}=
\nonumber\\
 & \frac{(m_s+m_b)^2}{4\pi^2}g_V^{(sb)}=\frac{(m_c+m_b)^2}{4\pi^2}g_V^{(cb)}=
\frac{m_b^2}{\pi^2}g_V^{(bb)}
\, ,\end{eqnarray}

\begin{eqnarray}
\label{7}
\Lambda=\frac{4\Lambda(ik)}
{(m_i+m_k)^2}.
\end{eqnarray}

\noindent
Here $m_i$ and $m_k$ are the quark masses in the intermediate state of
the quark loop. Dimensionless parameters $g$ and $\Lambda$ are supposed
to be constants which are independent of the quark interaction type. The
applicability of Eq. (\ref{5}) is verified by the success of
De Rujula-Georgi-Glashow quark model \cite{13}. Only the short-range
part of Breit potential connected with the gluon exchange is responsible
for the mass splitting in hadron multiplets.

We use the energy cutoff $\Lambda(ik)$ in the integrals
to avoid divergences.
The equation corresponding to Fig. 2 can be
written in the form:

\begin{eqnarray}
\label{8}
A_1(s,s_{12})&
=&\frac{\lambda_1 B_1(s_{12})}{1-B_1(s_{12})}\nonumber\\
&&\nonumber\\
&+&\frac{G_1(s_{12})}{1-B_1(s_{12})}
\int\limits_{4m^2}^{\Lambda(12)}
\, \frac{ds'_{12}}{\pi}\frac{\rho_1(s'_{12})}
{s'_{12}-s_{12}}G_1(s'_{12})\nonumber\\
&&\nonumber\\
 & \times & \int\limits_{-1}^{+1} \, \frac{dz}{2}
[A_1(s,s'_{13})+A_1(s,s'_{23})] \, .
\end{eqnarray}

In Eq. (\ref{8}) $z$ is the cosine of the angle between the relative momentum
of particles 1 and 2 in the intermediate state and the momentum of the third
particle in the final state in the c.m.s. of the particles 1 and 2. In
our case of equal mass of the quarks 1, 2 and 3, $s'_{13}$ and $s'_{12}$
are related by the equation (\ref{9})

\begin{eqnarray}
\label{9}
s'_{13}&=&2m^2-\frac{(s'_{12}+m^2-s)}{2}\\
&&\nonumber\\
\nonumber
 & \pm & \frac{z}{2} \sqrt{\frac{(s'_{12}-4m^2)}{s'_{12}}
(s'_{12}-(\sqrt{s}+m)^2)(s'_{12}-(\sqrt{s}-m)^2)}\, .
\end{eqnarray}

The expression for $s'_{23}$ is similar to (\ref{9}) with the replacement
$z\to -z$. This makes it possible to replace
$[A_1(s,s'_{13})+A_1(s,s'_{23})]$ in (\ref{8}) by $2A_1(s,s'_{13})$.

We extract the two-particle singularity from the amplitude $A_1(s,s_{12})$.
This singularity is the pole corresponding to the final diquark and is not
interesting for us:

\begin{eqnarray}
\label{10}
A_1(s,s_{12})=
\frac{\alpha_1(s,s_{12}) B_1(s_{12})}{1-B_1(s_{12})} \, .
\end{eqnarray}

The singularity corresponding to a bound baryon state
is contained in the reduced amplitude $\alpha_1(s,s_{12})$.

The equation for the reduced amplitude $\alpha_1(s,s_{12})$ can be written as

\begin{eqnarray}
\label{11}
\alpha_1(s,s_{12})&=&\lambda+\frac{1}{B_1(s_{12})}
\int\limits_{4m^2}^{\Lambda(12)}
\, \frac{ds'_{12}}{\pi}\frac{\rho_1(s'_{12})}
{s'_{12}-s_{12}}G_1(s'_{12})
\nonumber\\
&&\nonumber\\
& \times &
\int\limits_{-1}^{+1} \, \frac{dz}{2}
\, \frac{2\alpha_1(s,s'_{13}) B_1(s'_{13})}{1-B_1(s'_{13})} \, .
\end{eqnarray}

We construct the approximate solution of Eq. (\ref{11}).
We take into account only two-particle and three-particle singularities
and neglect others weaker ones.

The functions $G(s_{12})$ are practically independent of the
two-particle energy and we consider them equal to constants.
The function $\alpha_1(s,s_{12})$ is weakly dependent on the
two-particle energy.
For fixed values of $s$ and $s'_{12}$ the integration is carried out
over the region of the variable $s'_{13}$ corresponding to a physical
transition of the current into three quarks (the physical region of
Dalitz plot). It is convenient to take the central point of this region,
corresponding to $z=0$, to determinate the function $\alpha_1(s,s_{12})$
and also the Chew-Mandelstam function $B_1(s_{12})$ at the point
$s_{12}=s_0=\frac{s}{3}+m_b^2$. Then the equation for the $\Omega_{bbb}$
takes the form:

\begin{equation}
\label{12}
\alpha_1(s,s_0)=\lambda+I_{1,1}(s,s_0)\cdot 2\, \alpha_1(s,s_0)
\, , \end{equation}

\begin{equation}
\label{13}
I_{1,1}(s,s_0)=\int\limits_{4m^2_b}^{\Lambda_1}
\, \frac{ds'_{12}}{\pi} \frac{\rho_1(s'_{12})}
{s'_{12}-s_{12}}G_1\int\limits_{-1}^{+1} \, \frac{dz}{2}
\, \frac{G_1}{1-B_1(s'_{13})}
\, . \end{equation}

We can obtain an approximate solution of Eq. (\ref{13}):

\begin{equation}
\label{14}
\alpha_1(s,s_0)=\lambda [1-2\, I_{1,1}(s,s_0)]^{-1}
\, . \end{equation}

The pole in $s$ corresponds to a bound state of the baryon.

In analogy with the case of the $\Omega_{bbb}$ we can obtain the
rescattering amplitudes for all $S$-wave bottom baryons with $J^P=\frac{1}{2}^+$,
$\frac{3}{2}^+$, which include quarks of various flavours. These amplitudes will
satisfy systems of integral equations. Then we also solve them by approximate
method and find masses of the all $S$-wave bottom baryons multiplet.

In our calculation we use the following parameters of model: from the previous
paper \cite{15} quark masses are given: $m_{u,d}=495\, MeV$, $m_s=770\, MeV$, and
$m_c=1655\, MeV$; cutoff parameters $\lambda_q=10.7$ ($q=u, d, s$), $\lambda_c=6.5$;
gluon coupling constants $g_0=0.70$, $g_1=0.55$ for $J^p=0^+$ and $1^+$ light
diquarks, $g_c=0.857$ for charmed diquarks. As usual we have
$\lambda_{qQ}=\frac{1}{4}(\sqrt{\lambda_q}+\sqrt{\lambda_Q})^2$
($q=u, d, s$, $Q=c,b$). We use only two new parameters
$\lambda_b=5.4$ and $g_b=1.03$. These values have been determined by the $b$-baryon
masses: $M_{\Sigma_b \frac{1}{2}^+}=5.808\, GeV$ and $M_{\Sigma_b \frac{3}{2}^+}=5.829\, GeV$.
In order to fix $m_b=4.840\, GeV$ we use the $b$-baryon masses
$M_{\Sigma_b \frac{3}{2}^+}=5.829\, GeV$.

\section{The wave function of ${\bf (70,1^-)}$ excited bottom states.}
We consider a three-quark system having one unit of orbital
excitation. We take into account $u$, $d$, $b$-quarks. The orbital part of
wave function must have a mixed symmetry. The spin-flavor part of wave
function possesses the same symmetry in order to obtain a totally
symmetric state in the orbital-spin-flavor space.

As an example we derived the wave functions for the decuplets
$(10,2)$. The fully symmetric wave function for the decuplet state can be
constructed as:

\begin{eqnarray}
\label{15}
\varphi=\frac{1}{\sqrt{2}}\left(
\varphi_{MA}^{SU(6)}\varphi_{MA}^{O(3)}+
\varphi_{MS}^{SU(6)}\varphi_{MS}^{O(3)}
\right).
\end{eqnarray}

Then we obtain:

\begin{eqnarray}
\label{16}
\varphi=\frac{1}{\sqrt{2}}\varphi_{S}^{SU(3)}\left(
\varphi_{MA}^{SU(2)}\varphi_{MA}^{O(3)}+
\varphi_{MS}^{SU(2)}\varphi_{MS}^{O(3)}
\right),
\end{eqnarray}

\noindent
here $MA$ and $MS$ define the mixed antisymmetric and symmetric part
of wave function,

\begin{eqnarray}
\label{17}
\varphi_{MA}^{SU(6)}=\varphi_{S}^{SU(3)}\varphi_{MA}^{SU(2)},\quad\quad
\varphi_{MS}^{SU(6)}=\varphi_{S}^{SU(3)}\varphi_{MS}^{SU(2)}.
\end{eqnarray}

The functions $\varphi_{MA}^{SU(2)}$, $\varphi_{MS}^{SU(2)}$,
$\varphi_{MA}^{O(3)}$, $\varphi_{MS}^{O(3)}$ are given:

\begin{eqnarray}
\label{18}
\varphi_{MA}^{SU(2)}=\frac{1}{\sqrt{2}}\left(
\uparrow \downarrow \uparrow-\downarrow \uparrow \uparrow
\right),\quad\quad
\varphi_{MS}^{SU(2)}=\frac{1}{\sqrt{6}}\left(
\uparrow \downarrow \uparrow+\downarrow \uparrow \uparrow-
2\uparrow \uparrow \downarrow
\right),
\end{eqnarray}

\begin{eqnarray}
\label{19}
\varphi_{MA}^{O(3)}=\frac{1}{\sqrt{2}}\left(
010-100
\right),\quad\quad
\varphi_{MS}^{O(3)}=\frac{1}{\sqrt{6}}\left(
010+100-2\cdot 001
\right).
\end{eqnarray}

$\uparrow$ and $\downarrow$ determine the spin directions. $1$ and $0$
correspond to the excited or nonexcited quarks. The three projections
of orbital angular momentum are $l_z=1, 0, -1$. The $(10,2)$ multiplet with
$J^p=\frac{3}{2} ^{-}$ can be obtained using the spin $S=\frac{1}{2}$
and $l_z=1$, but the $(10,2)$ multiplet with $J^p=\frac{1}{2} ^{-}$
is determined by the spin $S=\frac{1}{2}$ and $l_z=0$.

We construct the $SU(3)$-function for each particle of multiplet.
For instance, the $SU(3)$-function for $\Sigma^{+}_b$-hyperon of decuplet
have following form:

\begin{eqnarray}
\label{20}
\varphi_{S}^{SU(3)}=\frac{1}{\sqrt{3}}\left(
ubu+buu+uub
\right).
\end{eqnarray}

We obtain the $SU(6)\times O(3)$-function for the $\Sigma^{+}_b$ of
the $(10,2)$ multiplet:

\begin{eqnarray}
\label{21}
\varphi_{\Sigma^{+}_b(10,2)}=\frac{\sqrt{6}}{18}\left(
2\{u^1\downarrow u\uparrow b\uparrow\}+
\{b^1\downarrow u\uparrow u\uparrow\}-\right.
\nonumber \\
\left.-\{u^1\uparrow u\downarrow b\uparrow\}-
\{u^1\uparrow u\uparrow b\downarrow \}-
\{b^1\uparrow u\uparrow u\downarrow \}\right).
\end{eqnarray}

Here the parenthesis determine the symmetrical function:

\begin{eqnarray}
\label{22}
\{{abc}\}\equiv abc+acb+bac+cab+bca+cba.
\end{eqnarray}

The wave functions of $\Sigma^{0}_b$- and $\Sigma^{-}_b$-hyperons can be
constructed by similar way.

For the $\Xi^{0,-}_{bb}$ state of the $(10,2)$ multiplet the wave function
is similar to the $\Sigma^{+,-}_b$ state with the replacement by
$u\leftrightarrow b$ or $d\leftrightarrow b$. The wave function for the
$\Omega_{bbb}$ of $(10,2)$ decuplet is determined as:

\begin{eqnarray}
\label{23}
\varphi_{\Omega_{bbb}(10,2)}=\frac{\sqrt{2}}{6}\left(
\{b^1\downarrow b\uparrow b\uparrow\}
-\{b^1\uparrow b\uparrow b\downarrow\}
\right).
\end{eqnarray}

The wave functions and the method of the construction for the multiplets
$(8,2)$, $(8,4)$ and $(1,2)$ are similar.

\section{The three-quark integral equations for the ${\bf (70,1^-)}$ bottom
multiplet.}

By the construction of $(70,1^-)$ bottom baryon multiplet integral
equations we need to using the projectors for the different diquark states.
The projectors to the symmetric and antisymmetric states can be obtained as:

\begin{eqnarray}
\label{24}
\frac{1}{2}\left(q_1 q_2+q_2 q_1\right),\quad\quad
\frac{1}{2}\left(q_1 q_2-q_2 q_1\right).
\end{eqnarray}

\begin{eqnarray}
\label{25}
\frac{1}{2}\left(\uparrow\downarrow+\downarrow\uparrow
\right),\quad\quad
\frac{1}{2}\left(\uparrow\downarrow-\downarrow\uparrow
\right).
\end{eqnarray}

\begin{eqnarray}
\label{26}
\frac{1}{2}\left(10+01\right),\quad\quad
\frac{1}{2}\left(10-01\right).
\end{eqnarray}

One can obtain the four types of totally symmetric projectors:

\begin{eqnarray}
\label{27}
S=S\cdot S\cdot S=\frac{1}{8}\left(q_1 q_2+q_2 q_1\right)
\left(\uparrow\downarrow+\downarrow\uparrow\right)\left(10+01\right),
\end{eqnarray}

\begin{eqnarray}
\label{28}
S=S\cdot A\cdot A=\frac{1}{8}\left(q_1 q_2+q_2 q_1\right)
\left(\uparrow\downarrow-\downarrow\uparrow\right)\left(10-01\right),
\end{eqnarray}

\begin{eqnarray}
\label{29}
S=A\cdot A\cdot S=\frac{1}{8}\left(q_1 q_2-q_2 q_1\right)
\left(\uparrow\downarrow-\downarrow\uparrow\right)\left(10+01\right),
\end{eqnarray}

\begin{eqnarray}
\label{30}
S=A\cdot S\cdot A=\frac{1}{8}\left(q_1 q_2-q_2 q_1\right)
\left(\uparrow\downarrow+\downarrow\uparrow\right)\left(10-01\right).
\end{eqnarray}

We use these projectors for the consideration of various diquarks:

\begin{eqnarray}
u^1\uparrow b\downarrow\,\,: \nonumber
\end{eqnarray}

\begin{eqnarray}
\label{31}
\frac{A^{0b}_{1}}{8}
\left(u^1\uparrow b\downarrow+u^1\downarrow b\uparrow+
b^1\uparrow u\downarrow+b^1\downarrow u\uparrow+
u\uparrow b^1\downarrow+u\downarrow b^1\uparrow+
b\uparrow u^1\downarrow+b\downarrow u^1\uparrow \right) \nonumber\\
+\frac{A^{1b}_{0}}{8}
\left(u^1\uparrow b\downarrow-u^1\downarrow b\uparrow+
b^1\uparrow u\downarrow-b^1\downarrow u\uparrow-
u\uparrow b^1\downarrow+u\downarrow b^1\uparrow-
b\uparrow u^1\downarrow+b\downarrow u^1\uparrow \right) \nonumber\\
+\frac{A^{0b}_{0}}{8}
\left(u^1\uparrow b\downarrow-u^1\downarrow b\uparrow-
b^1\uparrow u\downarrow+b^1\downarrow u\uparrow+
u\uparrow b^1\downarrow-u\downarrow b^1\uparrow-
b\uparrow u^1\downarrow+b\downarrow u^1\uparrow \right) \nonumber\\
+\frac{A^{1b}_{1}}{8}
\left(u^1\uparrow b\downarrow+u^1\downarrow b\uparrow-
b^1\uparrow u\downarrow-b^1\downarrow u\uparrow-
u\uparrow b^1\downarrow-u\downarrow b^1\uparrow+
b\uparrow u^1\downarrow+b\downarrow u^1\uparrow \right),
\end{eqnarray}

\begin{eqnarray}
u^1\uparrow b\uparrow\,\, : \nonumber
\end{eqnarray}

\begin{eqnarray}
\label{32}
\frac{A^{0b}_{1}}{4}
\left(u^1\uparrow b\uparrow+b^1\uparrow u\uparrow+
u\uparrow b^1\uparrow+b\uparrow u^1\uparrow\right)+ \nonumber \\
+\frac{A^{1b}_{1}}{4}
\left(u^1\uparrow b\uparrow-b^1\uparrow u\uparrow-
u\uparrow b^1\uparrow+b\uparrow u^1\uparrow
\right),
\end{eqnarray}

\begin{eqnarray}
u\uparrow b\downarrow\,\, : \nonumber
\end{eqnarray}

\begin{eqnarray}
\label{33}
\frac{A^{0b}_{1}}{4}
\left(u\uparrow b\downarrow+u\downarrow b\uparrow+
b\uparrow u\downarrow+b\downarrow u\uparrow
\right)+ \nonumber \\
+\frac{A^{0b}_{0}}{4}
\left(u\uparrow b\downarrow-u\downarrow b\uparrow-
b\uparrow u\downarrow+b\downarrow u\uparrow
\right),
\end{eqnarray}

\begin{eqnarray}
u\uparrow b\uparrow\,\, : \nonumber
\end{eqnarray}

\begin{eqnarray}
\label{34}
\frac{A^{0b}_{1}}{2}
\left(u\uparrow b\uparrow+b\uparrow u\uparrow
\right).
\end{eqnarray}

Here the lower index determines the value of spin projection,
and the upper index corresponds to the value of orbital angular momentum.

We consider the projectors (\ref{35}) - (\ref{38}), which are similar to (\ref{31}) - (\ref{34}) with
the replacement by $b\rightarrow u$ and use the amplitudes
$A^{0}_{1}$, $A^{1}_{0}$, $A^{0}_{0}$, $A^{1}_{1}$. The $A$ is the
three-quark amplitude.

\begin{eqnarray}
u^1\uparrow u\downarrow\,\, : \nonumber
\end{eqnarray}

\begin{eqnarray}
\label{35}
\frac{A^{0}_{1}}{4}
\left(u^1\uparrow u\downarrow+u^1\downarrow u\uparrow+
u\uparrow u^1\downarrow+u\downarrow u^1\uparrow
\right)+\nonumber\\
+\frac{A^{1}_{0}}{4}
\left(u^1\uparrow u\downarrow-u^1\downarrow u\uparrow-
u\uparrow u^1\downarrow+u\downarrow u^1\uparrow
\right),
\end{eqnarray}

\begin{eqnarray}
u^1\uparrow u\uparrow\,\, : \nonumber
\end{eqnarray}

\begin{eqnarray}
\label{36}
\frac{A^{0}_{1}}{2}
\left(u^1\uparrow u\uparrow+u\uparrow u^1\uparrow
\right),\end{eqnarray}

\begin{eqnarray}
u\uparrow u\downarrow\,\, : \nonumber
\end{eqnarray}

\begin{eqnarray}
\label{37}
\frac{A^{0}_{1}}{2}
\left(u\uparrow u\downarrow+u\downarrow u\uparrow
\right),
\end{eqnarray}

\begin{eqnarray}
u\uparrow u\uparrow\,\, : \nonumber
\end{eqnarray}

\begin{eqnarray}
\label{38}
A^{0}_{1}\, u\uparrow u\uparrow .
\end{eqnarray}

Here we consider the projection of orbital angular momentum $l_z=+1$. We
use only diquarks $1^+$, $0^+$, $2^-$, $1^-$. If we consider the $l_z=-1$
or $l_z=0$, that we obtain the other diquarks: $1^+$, $0^+$, $1^-$, $0^-$.
In our model the five types of diquarks $1^+$, $0^+$, $2^-$, $1^-$, $0^-$
are constructed.

We use the diquark projectors and consider the particle $\Sigma_b$
$\frac{3}{2} ^{-}$ of the $(10,2)$ multiplet. This wave function
contains the contribution $u^1\downarrow u\uparrow b\uparrow$, which
includes three diquarks: $u^1\downarrow u\uparrow$,\,
$u^1\downarrow b\uparrow$\, and \, $u\uparrow b\uparrow$.
The diquark projectors allow us to obtain the equations (\ref{39}) -- (\ref{41})
(with the definition $\rho_J(s_{ij})\equiv k_{ij}$).

\begin{eqnarray}
\label{39}
k_{12}\left(\frac{A_1^0+A_0^1}{4}\left(u^1\downarrow u\uparrow b\uparrow
+u\uparrow u^1\downarrow b\uparrow\right)+\right. \nonumber\\
\left. +\frac{A_1^0-A_0^1}{4}\left(
u^1\uparrow u\downarrow b\uparrow+u\downarrow u^1\uparrow b\uparrow
\right)\right)\, ,
\end{eqnarray}

\begin{eqnarray}
\label{40}
k_{13}\left(\frac{A_1^{0b}+A_0^{1b}+A_0^{0b}+A_1^{1b}}{8}
\left(u^1\downarrow u\uparrow b\uparrow
+b\uparrow u\uparrow u^1\downarrow\right)+\right. \nonumber\\
\left. +\frac{A_1^{0b}-A_0^{1b}-A_0^{0b}+A_1^{1b}}{8}
\left(u^1\uparrow u\uparrow b\downarrow
+b\downarrow u\uparrow u^1\uparrow\right)+\right. \nonumber\\
\left. +\frac{A_1^{0b}+A_0^{1b}-A_0^{0b}-A_1^{1b}}{8}
\left(b^1\downarrow u\uparrow u\uparrow
+u\uparrow u\uparrow b^1\downarrow\right)+\right. \nonumber\\
\left. +\frac{A_1^{0b}-A_0^{1b}+A_0^{0b}-A_1^{1b}}{8}
\left(b^1\uparrow u\uparrow u\downarrow
+u\downarrow u\uparrow b^1\uparrow\right)
\right)\, ,
\end{eqnarray}

\begin{eqnarray}
\label{41}
k_{23}\left(\frac{A_1^{0b}}{2}\left(u^1\downarrow u\uparrow b\uparrow
+u^1\downarrow b\uparrow u\uparrow\right)
\right)\, .
\end{eqnarray}

Then all members of wave function can be considered. After the grouping of
these members we can obtain:

\begin{eqnarray}
\label{42}
u^1\downarrow u\uparrow b\uparrow \left\{
k_{12}\frac{A_1^0+3A_0^1}{4}+k_{13}\frac{A_1^{0b}+3A_0^{1b}}{4}
+k_{23}\, A_1^{0b}\right\}\, .
\end{eqnarray}

We group the same members and obtain the system integral equations
for the $\Sigma_b$ state with the $J^p=\frac{3}{2} ^{-}$ $(10,2)$
multiplet:

\begin{eqnarray}
\label{43}
A_1^0(s,s_{12})&=&\lambda\, b_{1^+}(s_{12})L_{1^+}(s_{12})+
K_{1^+}(s_{12})\left[\frac{1}{4}A_1^{0b}(s,s_{13})+
\frac{3}{4}A_0^{1b}(s,s_{13})+\right.\nonumber\\
&&\nonumber\\
&+&\left.\frac{1}{4}A_1^{0b}(s,s_{23})+\frac{3}{4}A_0^{1b}(s,s_{23})
\right]\\
\label{44}
A_1^{0b}(s,s_{13})&=&\lambda\, b_{1_b^+}(s_{13})L_{1_b^+}(s_{13})+
K_{1_b^+}(s_{13})\left[\frac{1}{2}A_1^0(s,s_{12})-
\frac{1}{4}A_1^{0b}(s,s_{12})+\right.\nonumber\\
&&\nonumber\\
&+&\left.\frac{3}{4}A_0^{1b}(s,s_{12})+\frac{1}{2}A_1^0(s,s_{23})-
\frac{1}{4}A_1^{0b}(s,s_{23})+\frac{3}{4}A_0^{1b}(s,s_{23})
\right]\\
&&\nonumber\\
\label{45}
A_0^{1b}(s,s_{23})&=&\lambda\, b_{1_b^-}(s_{23})L_{1_b^-}(s_{23})+
K_{1_b^-}(s_{23})\left[\frac{1}{2}A_1^0(s,s_{12})+
\frac{1}{4}A_1^{0b}(s,s_{12})+\right.\nonumber\\
&&\nonumber\\
&+&\left.\frac{1}{4}A_0^{1b}(s,s_{12})+\frac{1}{2}A_1^0(s,s_{13})+
\frac{1}{4}A_1^{0b}(s,s_{13})+\frac{1}{4}A_0^{1b}(s,s_{13})
\right] \, .
\end{eqnarray}

Here function $L_J(s_{ik})$ has the form

\begin{eqnarray}
\label{46}
L_J(s_{ik})=\frac{G_J(s_{ik})}{1-B_J(s_{ik})}.
\end{eqnarray}

The integral operator $K_J (s_{ik})$ is:

\begin{eqnarray}
\label{47}
K_J (s_{ik})=L_J(s_{ik})\, \int\limits_{(m_i+m_k)^2}^{\Lambda_J \frac{(m_i+m_k)^2}{4}}\,
\frac{ds'_{ik}}{\pi}\frac{\rho_J(s'_{ik})G_J(s'_{ik})}
{s'_{ik}-s_{ik}}\, \int\limits_{-1}^{1}\frac{dz}{2}\, .
\end{eqnarray}

\begin{eqnarray}
\label{48}
b_J(s_{ik})=\int\limits_{(m_i+m_k)^2}^{\Lambda_J \frac{(m_i+m_k)^2}{4}}\,
\frac{ds'_{ik}}{\pi}\frac{\rho_J(s'_{ik})G_J(s'_{ik})}
{s'_{ik}-s_{ik}}\, .
\end{eqnarray}

The function $b_J(s_{ik})$ is the truncated function of Chew-Mandelstam.
$z$ is the cosine of the angle between the relative momentum of particles
$i$ and $k$ in the intermediate state and the momentum of
particle $j$ in the final state, taken in the c.m. of the particles
$i$ and $k$. Let some current produces three quarks (first diagram Fig.1)
with the vertex constant $\lambda$. This constant do not affect to the
spectra mass of excited baryons.

By analogy with the $\Sigma_b$ $\frac{3}{2} ^{-}$ $(10,2)$ state we obtain
the rescattering amplitudes of the three various quarks for all $P$-wave
states of the $(70,1^-)$ multiplet which satisfy the system of
integral equations.

\section{The reduced equations of ${\bf (70,1^-)}$ multiplet.}

Let us extract two-particle singularities in $A_J(s,s_{ik})$:

\begin{eqnarray}
\label{49}
A_J(s,s_{ik})=\frac{\alpha_J(s,s_{ik})b_J(s_{ik})G_J(s_{ik})}
{1-B_J(s_{ik})},
\end{eqnarray}

\noindent
$\alpha_J(s,s_{ik})$ is the reduced amplitude. Accordingly all integral
equations can be rewritten using the reduced amplitudes. For instance, one
consider the first equation of system for the $\Sigma_b$ $J^p=\frac{3}{2}^-$
of the $(10,2)$ multiplet:

\begin{eqnarray}
\label{50}
\alpha_1^0 (s,s_{12})&=&\lambda+\frac{1}{b_{1^+}(s_{12})}
\, \int\limits_{(m_1+m_2)^2}^{\Lambda_{1^+}(1,2)}\,
\frac{ds'_{12}}{\pi}\,\frac{\rho_{1^+}(s'_{12})G_{1^+}(s'_{12})}
{s'_{12}-s_{12}}\nonumber\\
&\times&\int\limits_{-1}^{1}\frac{dz}{2}\,
\left(
\frac{G_{1_b^+}(s'_{13})b_{1_b^+}(s'_{13})}{1-B_{1_b^+}(s'_{13})}
\,\frac{1}{2}\,\alpha_1^{0b}(s,s'_{13})+
\frac{G_{1_b^-}(s'_{13})b_{1_b^-}(s'_{13})}{1-B_{1_b^-}(s'_{13})}
\,\frac{3}{2}\,\alpha_0^{1b}(s,s'_{13})
\right).
\end{eqnarray}

The connection between $s'_{12}$ and $s'_{13}$ is:

\begin{eqnarray}
\label{51}
s'_{13}&=&m_1^2+m_3^2-\frac{\left(s'_{12}+m_3^2-s\right)
\left(s'_{12}+m_1^2-m_2^2\right)}{2s'_{12}}\nonumber\\
&&\nonumber\\
&\pm &\frac{z}{2s'_{12}}\times\sqrt{\left(s'_{12}-(m_1+m_2)^2\right)
\left(s'_{12}-(m_1-m_2)^2\right)}\nonumber\\
&&\nonumber\\
&\times&\sqrt{\left(s'_{12}-(\sqrt{s}+m_3)^2\right)
\left(s'_{12}-(\sqrt{s}-m_3)^2\right)}\, .
\end{eqnarray}

The formula for $s'_{23}$ is similar to (\ref{51}) with $z$ replaced by $-z$.
Thus $A_1^{0b}(s,s'_{13})+A_1^{0b}(s,s'_{23})$ must be replaced by
$2A_1^{0b}(s,s'_{13})$. $\Lambda_J(i,k)$ is the cutoff at the large
value of $s_{ik}$, which determines the contribution from small distances.

\begin{table}
\caption{The $\Lambda_b$-hyperon masses of multiplet $(70,1^-)$.}\label{tab1}
\begin{tabular}{cccc}
\toprule[1pt]
Multiplet & Baryon & Mass ($GeV$) & Mass ($GeV$) (exp.) \\
\midrule[1pt]
$\frac{5}{2}^-$ $(8,4)$ & $D_{05}$ & 6027 & -- \\
$\frac{3}{2}^-$ $(8,4)$ & $D_{03}$ & 5900 &  -- \\
$\frac{1}{2}^-$ $(8,4)$ & $S_{01}$ & 6159 & -- \\
$\frac{3}{2}^-$ $(8,2)$ & $D_{03}$ & 5920 & 5920 \\
$\frac{1}{2}^-$ $(8,2)$ & $S_{01}$ & 5912 & 5912 \\
$\frac{3}{2}^-$ $(1,2)$ & $D_{03}$ & -- & -- \\
$\frac{1}{2}^-$ $(1,2)$ & $S_{01}$ & -- & -- \\
\bottomrule[1pt]
\end{tabular}
\end{table}

The construction of the approximate solution of the (\ref{43}) -- (\ref{45})
is based on the extraction of the leading singularities which are close to the
region $s_{ik}=(m_i+m_k)^2$. Amplitudes with different number of
rescattering have the following structure of singularities. The main
singularities in $s_{ik}$ are from pair rescattering of the particles
$i$ and $k$. First of all there are threshold square root singularities.
Also possible are pole singularities, which correspond to the bound
states. The diagrams in Fig.2 apart from two-particle singularities
have their own specific triangle singularities. Such classification
allows us to search the approximate solution of (\ref{43}) -- (\ref{45}) by taking into
account some definite number of leading singularities and neglecting all
the weaker ones.

We consider the approximation, which corresponds to the single interaction
of all three particles (two-particle and triangle singularities) and
neglecting all the weaker ones.

The functions $\alpha_J(s,s_{ik})$ are the smooth functions of $s_{ik}$
as compared with the singular part of the amplitude, hence it can be
expanded in a series in the singulary point and only the first term of
this series should be employed further. As $s_0$ it is convenient to
take the middle point of physical region of Dalitz-plot in which $z=0$.
In this case we get
$s_{ik}=s_0=\frac{s+m_1^2+m_2^2+m_3^2}{m_{12}^2+m_{13}^2+m_{23}^2}$,
where $m_{ik}=\frac{m_i+m_k}{2}$. We define $\alpha_J(s,s_{ik})$ and
$b_J(s_{ik})$ at the point $s_0$. Such a choice of point $s_0$ allows us
to replace integral equations (\ref{43}) -- (\ref{45}) by the algebraic equations for the
state $\Sigma_b$ $J^p=\frac{3}{2}^-$ of the $(10,2)$ multiplet:

\begin{eqnarray}
\label{52}
\alpha_1^0(s,s_0)&=&\lambda+\frac{1}{2}\,\alpha_1^{0b}(s,s_0)
\, I_{1^+ 1^+_b}(s,s_0)\,\frac{b_{1^+_b}(s_0)}{b_{1^+}(s_0)}
+\frac{3}{2}\,\alpha_0^{1b}(s,s_0)\, I_{1^+ 1^-_b}(s,s_0)
\,\frac{b_{1^-_b}(s_0)}{b_{1^+}(s_0)}
\hskip5.5ex 1^+\\
&& \nonumber\\
\label{53}
\alpha_1^{0b}(s,s_0)&=&\lambda+\alpha_1^0(s,s_0)\, I_{1^+_b 1^+}(s,s_0)
\,\frac{b_{1^+}(s_0)}{b_{1^+_b}(s_0)}-\frac{1}{2}\,\alpha_1^{0b}(s,s_0)
\, I_{1^+_b 1^+_b}(s,s_0)\hskip17ex 1^+_b \nonumber\\
&& \nonumber\\
&+&\frac{3}{2}\,\alpha_0^{1b}(s,s_0)\, I_{1^+_b 1^-_b}(s,s_0)
\,\frac{b_{1^-_b}(s_0)}{b_{1^+_b}(s_0)}\\
&& \nonumber\\
\label{54}
\alpha_0^{1b}(s,s_0)&=&\lambda+\alpha_1^0(s,s_0)\, I_{1^-_b 1^+}(s,s_0)
\,\frac{b_{1^+}(s_0)}{b_{1^-_b}(s_0)}+\frac{1}{2}\,\alpha_1^{0b}(s,s_0)
\, I_{1^-_b 1^+_b}(s,s_0)\,\frac{b_{1^+_b}(s_0)}{b_{1^-_b}(s_0)}
\hskip8.8ex 1^-_b \nonumber\\
&& \nonumber\\
&+&\frac{1}{2}\,\alpha_0^{1b}(s,s_0)\, I_{1^-_b 1^-_b}(s,s_0)\, .
\end{eqnarray}

Here the reduced amplitudes for the diquarks $1^+$, $1^+_b$, $1^-_b$
are given. The function $I_{J_1 J_2}(s,s_0)$ takes into account singularity
which corresponds to the simultaneous vanishing of all propagators in the
triangle diagrams.

\begin{eqnarray}
\label{55}
I_{J_1 J_2}(s,s_0)=\int\limits_{(m_i+m_k)^2}^{\Lambda_J \frac{(m_i+m_k)^2}{4}}\,
\frac{ds'_{ik}}{\pi}\frac{\rho_{J_1}(s'_{ik})G^2_{J_1}(s'_{ik})}
{s'_{ik}-s_{ik}}\, \int\limits_{-1}^{1}\frac{dz}{2}\,
\frac{1}{1-B_{J_2}(s_{ij})}\, .
\end{eqnarray}

The $G_J(s_{ik})$ functions have the smooth dependence from energy
$s_{ik}$ therefore we suggest them as constants.

We calculate the system equations and can determine the mass values
of the $\Sigma_b$ $J^p=\frac{3}{2}^-$ $(10,2)$. We calculate a pole in $s$
which corresponds to the bound state of three quarks.

\begin{table}
\caption{The $\Sigma_b$-hyperon masses of multiplet $(70,1^-)$.}\label{tab2}
\begin{tabular}{cccc}
\toprule[1pt]
Multiplet & Baryon & Mass ($GeV$) & Mass ($GeV$) (exp.) \\
\midrule[1pt]
$\frac{3}{2}^-$ $(10,2)$ & $D_{13}$ & 4057 & --\\
$\frac{1}{2}^-$ $(10,2)$ & $S_{11}$ & 5220 & --\\
$\frac{5}{2}^-$ $(8,4)$ & $D_{15}$ & 4649 & --\\
$\frac{3}{2}^-$ $(8,4)$ & $D_{13}$ & 4057 & --\\
$\frac{1}{2}^-$ $(8,4)$ & $S_{11}$ & 5220 & --\\
$\frac{3}{2}^-$ $(8,2)$ & $D_{13}$ & 4161 & --\\
$\frac{1}{2}^-$ $(8,2)$ & $S_{11}$ & 4523 & --\\
\bottomrule[1pt]
\end{tabular}
\end{table}

By analogy with $\Sigma_b$-hyperon we obtain the system equations for the
reduced amplitudes for all particles $(70,1^-)$ multiplets.

\section{Calculation results.}

The present paper is an extension of previous research works devoted
to strange and charmed $P$-wave $(70,1^-)$ baryon multiplets. Therefore
we take the part of parameters from those ones. The mass of $u$, $d$ quarks
is equal to $m_{u,d}=570\, MeV$ as in those works, the mass of $b$ quark
is taken equal to $m_b=5085\, MeV$. As in the case of charmed $P$-wave
$(70,1^-)$ baryon, in the case of bottom $P$-wave baryons we have to
shift mass from their usual meanings by $245\, MeV$ ($\Delta=1900-1655=245=5085-4840$).
This shift takes into account the confinement potential and allows baryons
to remain below the threshold.

\begin{table}
\caption{The $\Xi_{bb}$-hyperon masses of multiplet $(70,1^-)$.}\label{tab3}
\begin{tabular}{cccc}
\toprule[1pt]
Multiplet & Baryon & Mass ($GeV$) & Mass ($GeV$) (exp.) \\
\midrule[1pt]
$\frac{3}{2}^-$ $(10,2)$ & $D_{13}$ & 9586 & --\\
$\frac{1}{2}^-$ $(10,2)$ & $S_{11}$ & 10394 & --\\
$\frac{5}{2}^-$ $(8,4)$ & $D_{15}$ & 9483 & --\\
$\frac{3}{2}^-$ $(8,4)$ & $D_{13}$ & 9100 & --\\
$\frac{1}{2}^-$ $(8,4)$ & $S_{11}$ & 10016 & --\\
$\frac{3}{2}^-$ $(8,2)$ & $D_{13}$ & 9130 & --\\
$\frac{1}{2}^-$ $(8,2)$ & $S_{11}$ & 9399 & --\\
\bottomrule[1pt]
\end{tabular}
\end{table}

\begin{table}
\caption{The $\Omega_{bbb}$-hyperon masses of multiplet $(70,1^-)$.}\label{tab4}
\begin{tabular}{cccc}
\toprule[1pt]
Multiplet & Baryon & Mass ($GeV$) & Mass ($GeV$) (exp.) \\
\midrule[1pt]
$\frac{3}{2}^-$ $(10,2)$ & $D_{03}$ & 14940 & --\\
$\frac{1}{2}^-$ $(10,2)$ & $S_{01}$ & 14984 & --\\
\bottomrule[1pt]
\end{tabular}
\end{table}

\begin{table}
\caption{Coefficient of Ghew-Mandelstam functions for the
different diquarks.}\label{tab5}
\begin{tabular}{cccc}
\toprule[1pt]
 &$\alpha_J$&$\beta_J$&$\delta_J$\\
\midrule[1pt]
 & & & \\
$1^+$&$\frac{1}{3}$&$\frac{4m_i m_k}{3(m_i+m_k)^2}-\frac{1}{6}$
&$-\frac{1}{6}(m_i-m_k)^2$\\
 & & & \\
$0^+$&$\frac{1}{2}$&$-\frac{1}{2}\frac{(m_i-m_k)^2}{(m_i+m_k)^2}$&0\\
 & & & \\
$0^-$&$0$&$\frac{1}{2}$&$-\frac{1}{2}(m_i-m_k)^2$\\
 & & & \\
$1^-$&$\frac{1}{2}$&$-\frac{1}{2}\frac{(m_i-m_k)^2}{(m_i+m_k)^2}$&0\\
 & & & \\
$2^-$&$\frac{3}{10}$&$\frac{1}{5}
\left(1-\frac{3}{2}\frac{(m_i-m_k)^2}{(m_i+m_k)^2}\right)$
&$-\frac{1}{5}(m_i-m_k)^2$\\
 & & & \\
\bottomrule[1pt]
\end{tabular}
\end{table}

The gluon coupling constants $g_+=0.69$, $g_-=0.3$ and cutoff $\Lambda_{uu}=14.5$
for the light quarks are the similar to ones in work \cite{11}.

There are only two experimentally known masses of bottom $P$-wave baryons:
$\Lambda_b$ $\frac{3}{2}^-$ $M=5920\, MeV$ and $\Lambda_b$ $\frac{1}{2}^-$ $M=5912\, MeV$.
We use them to define the parameters for bottom diquarks:
$g_{ub,bb}=1.24$, $\Lambda_{ub}=9.1$, $\Lambda_{bb}=4.82$
($\Lambda_{ub}=\frac{1}{4}(\sqrt{\Lambda_{uu}}+\sqrt{\Lambda_{bb}})^2$
as usual).

We calculated 21 different baryon masses of the $(70,1^-)$ bottom baryon multiplet.

\section{Conclusions.}
In this section we will give a brief explanation of some symbols.

In accordance with the theorem on the connection of spin and statistics
fermion states should have an antisymmetric wave functions. Three-quark
wave functions of $S$-wave baryons have a structure corresponding to the
symmetry $SU(3)_f\times SU(2)\times SU(3)_c$, where multipliers are gropes
of flavor, spin, and color symmetry respectively. The part of the structure corresponding
to the $SU(3)_c$ is antisymmetric, other part corresponding to the $SU(6)=SU(3)_f\times SU(2)$
is symmetric. Combining three fundamental representations of $SU(6)$ yields
$6\otimes 6\otimes 6=56\oplus 70\oplus 70\oplus 20$, here 56 is the multiplet
possessing symmetric wave function, 70 are the multiplets possessing mixed
symmetric and mixed antisymmetric wave functions, and 20 is the multiplet
possessing antisymmetric wave function. In the case of $S$-wave baryons we have
the 56 multiplet. If we consider $P$-wave baryons, that is with the orbital
excitatin $L=1$, then we will have two 70 multiplets because in the
according with the oscillator model we will have two $O(3)$ mixed symmetry
wave functions. So in the case of $L=1$ we have $SU(6)\times O(3)\times SU(3)_c$
and write ${\bf (70,1^-)}$. This multiplet has negative parity.

\begin{acknowledgments}
The work was carried with the support of the Russian Ministry of Education
(grant 2.1.1.68.26) and RFBR, Research Project No. 13-02-91154.
\end{acknowledgments}

\newpage

\vskip60pt
\begin{picture}(600,80)
\put(-10,40){\line(1,0){33}}
\put(-10,50){\line(1,0){28}}
\put(-10,60){\line(1,0){33}}
\put(19,46){\line(1,1){15}}
\put(22,41){\line(1,1){17}}
\put(27.5,38.5){\line(1,1){14}}
\put(41,56){\vector(2,1){28}}
\put(42.5,50){\vector(1,0){35}}
\put(41,44){\vector(2,-1){28}}
\put(30,50){\circle{25}}
\put(70,78){$1$}
\put(70,55){$2$}
\put(70,20){$3$}
\put(87,47){$=$}
\put(107,53){\line(1,0){28}}
\put(107,50){\line(1,0){28}}
\put(107,47){\line(1,0){28}}
\put(135,53){\vector(2,1){28}}
\put(135,50){\vector(1,0){35}}
\put(135,47){\vector(2,-1){28}}
\put(163,78){$1$}
\put(163,55){$2$}
\put(163,20){$3$}
\put(180,47){$+$}
\put(200,40){\line(1,0){33}}
\put(200,50){\line(1,0){28}}
\put(200,60){\line(1,0){33}}
\put(229,46){\line(1,1){15}}
\put(232,41){\line(1,1){17}}
\put(237.5,38.5){\line(1,1){14}}
\put(251,44){\vector(2,-1){28}}
\put(240,50){\circle{25}}
\put(268,54){\oval(33,33)[tl]}
\put(252,70){\oval(33,33)[br]}
\put(269,71){\vector(2,3){15}}
\put(269,71){\vector(2,-1){24}}
\put(287,95){$1$}
\put(295,65){$2$}
\put(280,20){$3$}
\put(300,47){$+$}
\put(140,0){{\large a)}}
\put(250,0){{\large b)}}
\end{picture}

\vskip60pt
\begin{picture}(600,60)
\put(90,47){$+$}
\put(110,40){\line(1,0){33}}
\put(110,50){\line(1,0){28}}
\put(110,60){\line(1,0){33}}
\put(139,46){\line(1,1){15}}
\put(142,41){\line(1,1){17}}
\put(147.5,38.5){\line(1,1){14}}
\put(161,44){\vector(2,-1){28}}
\put(150,50){\circle{25}}
\put(178,54){\oval(33,33)[tl]}
\put(162,70){\oval(33,33)[br]}
\put(179,71){\vector(2,3){15}}
\put(179,71){\vector(2,-1){24}}
\put(197,95){$1$}
\put(205,65){$3$}
\put(190,20){$2$}
\put(210,47){$+$}
\put(230,40){\line(1,0){33}}
\put(230,50){\line(1,0){28}}
\put(230,60){\line(1,0){33}}
\put(259,46){\line(1,1){15}}
\put(262,41){\line(1,1){17}}
\put(267.5,38.5){\line(1,1){14}}
\put(281,44){\vector(2,-1){28}}
\put(270,50){\circle{25}}
\put(298,54){\oval(33,33)[tl]}
\put(282,70){\oval(33,33)[br]}
\put(299,71){\vector(2,3){15}}
\put(299,71){\vector(2,-1){24}}
\put(317,95){$2$}
\put(325,65){$3$}
\put(310,20){$1$}
\put(160,0){{\large c)}}
\put(275,0){{\large d)}}
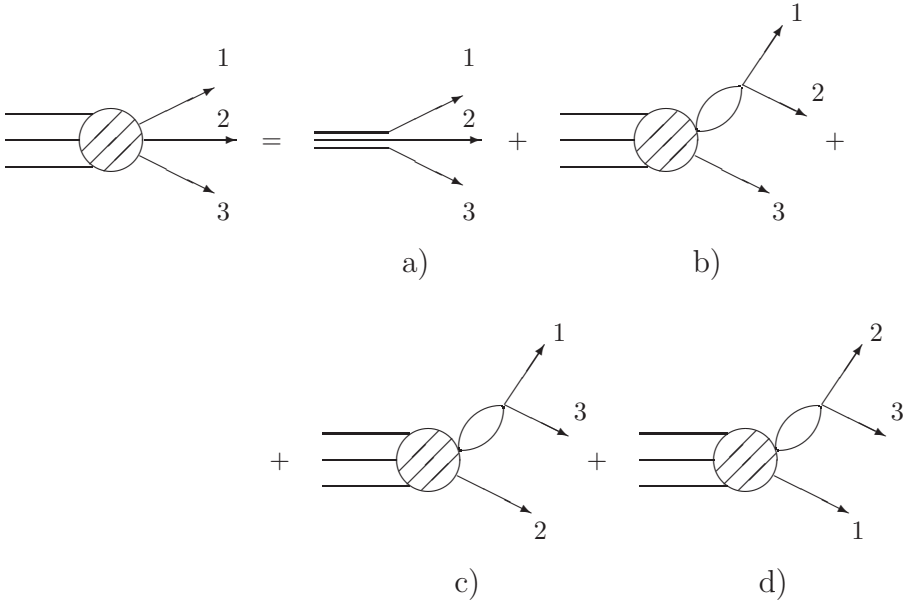
\put(-10,-30){{\large Fig.1. The contribution of diagrams at the last pair
of the interacting particles.}}
\end{picture}

\vskip150pt
\begin{picture}(600,60)
\put(-10,40){\line(1,0){33}}
\put(-10,50){\line(1,0){28}}
\put(-10,60){\line(1,0){33}}
\put(19,46){\line(1,1){15}}
\put(22,41){\line(1,1){17}}
\put(27.5,38.5){\line(1,1){14}}
\put(41,44){\vector(2,-1){28}}
\put(30,50){\circle{25}}
\put(58,54){\oval(33,33)[tl]}
\put(42,70){\oval(33,33)[br]}
\put(59,71){\vector(2,3){15}}
\put(59,71){\vector(2,-1){24}}
\put(77,95){$1$}
\put(85,65){$2$}
\put(70,20){$3$}
\put(90,47){$=$}
\put(110,52){\line(1,0){28}}
\put(110,50){\line(1,0){28}}
\put(110,48){\line(1,0){28}}
\put(154,51){\oval(33,33)[tl]}
\put(138,67){\oval(33,33)[br]}
\put(155,67){\vector(2,3){15}}
\put(155,67){\vector(2,-1){24}}
\put(139,49){\vector(2,-1){28}}
\put(173,91){$1$}
\put(181,61){$2$}
\put(168,25){$3$}
\put(190,47){$+$}
\put(210,40){\line(1,0){33}}
\put(210,50){\line(1,0){28}}
\put(210,60){\line(1,0){33}}
\put(239,46){\line(1,1){15}}
\put(242,41){\line(1,1){17}}
\put(247.5,38.5){\line(1,1){14}}
\put(261,44){\vector(1,0){43}}
\put(250,50){\circle{25}}
\put(278,54){\oval(33,33)[tl]}
\put(262,70){\oval(33,33)[br]}
\put(279,71){\vector(2,3){15}}
\put(279,71){\vector(1,-1){25}}
\put(297,95){$3$}
\put(300,60){$1$}
\put(290,27){$2$}
\put(305,29){\oval(33,33)[tr]}
\put(321,45){\oval(33,33)[bl]}
\put(323,29){\vector(2,3){15}}
\put(323,29){\vector(2,-1){24}}
\put(341,53){$1$}
\put(333,7){$2$}
\put(360,47){$+$}
\end{picture}

\vskip60pt
\begin{picture}(600,60)
\put(190,47){$+$}
\put(210,40){\line(1,0){33}}
\put(210,50){\line(1,0){28}}
\put(210,60){\line(1,0){33}}
\put(239,46){\line(1,1){15}}
\put(242,41){\line(1,1){17}}
\put(247.5,38.5){\line(1,1){14}}
\put(261,44){\vector(1,0){43}}
\put(250,50){\circle{25}}
\put(278,54){\oval(33,33)[tl]}
\put(262,70){\oval(33,33)[br]}
\put(279,71){\vector(2,3){15}}
\put(279,71){\vector(1,-1){25}}
\put(297,95){$3$}
\put(300,60){$2$}
\put(290,27){$1$}
\put(305,29){\oval(33,33)[tr]}
\put(321,45){\oval(33,33)[bl]}
\put(323,29){\vector(2,3){15}}
\put(323,29){\vector(2,-1){24}}
\put(341,53){$2$}
\put(333,7){$1$}
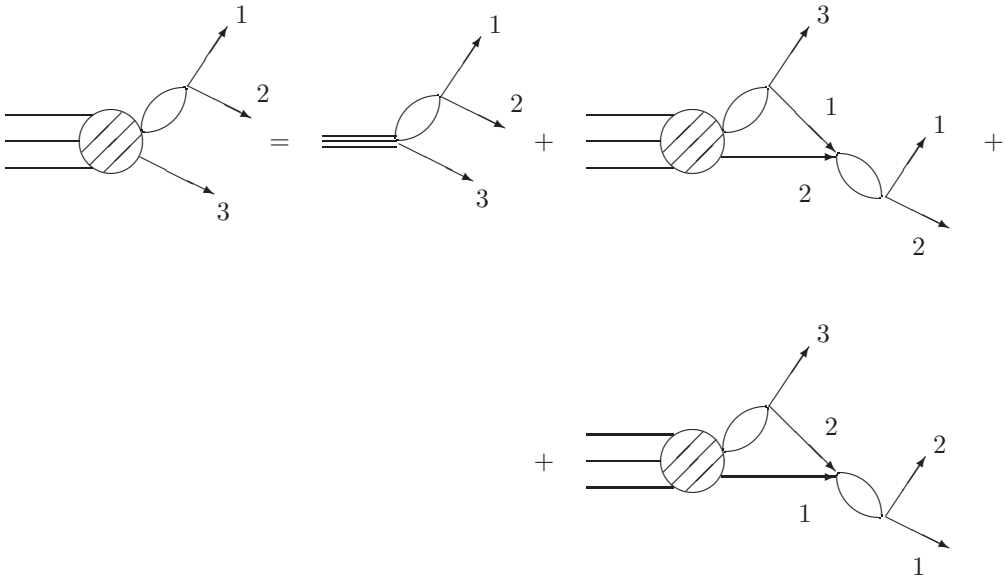
\put(-10,-10){{\large Fig.2. Graphic representation of the equations
for the amplitude $A_1(s,s_{ik})$.}}
\end{picture}

\vskip60pt
\begin{picture}(600,100)
\multiput(70,20)(0,5){20}{\line(0,1){2}}
\put(-10,50){\line(1,0){33}}
\put(-10,60){\line(1,0){28}}
\put(-10,70){\line(1,0){33}}
\put(19,56){\line(1,1){15}}
\put(22,51){\line(1,1){17}}
\put(27.5,48.5){\line(1,1){14}}
\put(41,54){\vector(1,0){43}}
\put(30,60){\circle{25}}
\put(58,64){\oval(33,33)[tl]}
\put(42,80){\oval(33,33)[br]}
\put(59,81){\vector(2,3){15}}
\put(59,81){\vector(1,-1){25}}
\put(77,105){$3$}
\put(75,70){$1, 2$}
\put(60,37){$2, 1$}
\put(90,90){$k_{13}, k_{23}$}
\put(85,39){\oval(33,33)[tr]}
\put(101,55){\oval(33,33)[bl]}
\put(103,39){\vector(2,3){15}}
\put(103,39){\vector(2,-1){24}}
\put(121,40){$k_{12}\frac{A_1^0+3A_0^1}{4}\left|_{k_{12}}\right.$}
\put(200,70){$+$}
\multiput(305,20)(0,5){20}{\line(0,1){2}}
\put(225,50){\line(1,0){33}}
\put(225,60){\line(1,0){28}}
\put(225,70){\line(1,0){33}}
\put(254,56){\line(1,1){15}}
\put(257,51){\line(1,1){17}}
\put(262.5,48.5){\line(1,1){14}}
\put(276,54){\vector(1,0){43}}
\put(265,60){\circle{25}}
\put(293,64){\oval(33,33)[tl]}
\put(277,80){\oval(33,33)[br]}
\put(294,81){\vector(2,3){15}}
\put(294,81){\vector(1,-1){25}}
\put(312,105){$2$}
\put(310,70){$1, 3$}
\put(295,37){$3, 1$}
\put(325,90){$k_{12}, k_{23}$}
\put(320,39){\oval(33,33)[tr]}
\put(336,55){\oval(33,33)[bl]}
\put(338,39){\vector(2,3){15}}
\put(338,39){\vector(2,-1){24}}
\put(356,40){$k_{13}\frac{A_1^{0b}+3A_0^{1b}}{4}\left|_{k_{13}}\right.$}
\put(430,70){$+$}
\end{picture}

\vskip60pt
\begin{picture}(600,80)
\multiput(170,20)(0,5){20}{\line(0,1){2}}
\put(65,70){$+$}
\put(90,50){\line(1,0){33}}
\put(90,60){\line(1,0){28}}
\put(90,70){\line(1,0){33}}
\put(119,56){\line(1,1){15}}
\put(122,51){\line(1,1){17}}
\put(127.5,48.5){\line(1,1){14}}
\put(141,54){\vector(1,0){43}}
\put(130,60){\circle{25}}
\put(158,64){\oval(33,33)[tl]}
\put(142,80){\oval(33,33)[br]}
\put(159,81){\vector(2,3){15}}
\put(159,81){\vector(1,-1){25}}
\put(177,105){$1$}
\put(175,70){$2, 3$}
\put(160,37){$3, 2$}
\put(190,90){$k_{12}, k_{13}$}
\put(185,39){\oval(33,33)[tr]}
\put(201,55){\oval(33,33)[bl]}
\put(203,39){\vector(2,3){15}}
\put(203,39){\vector(2,-1){24}}
\put(221,40){$k_{23}A_1^{0b}\left|_{k_{23}}\right.$}
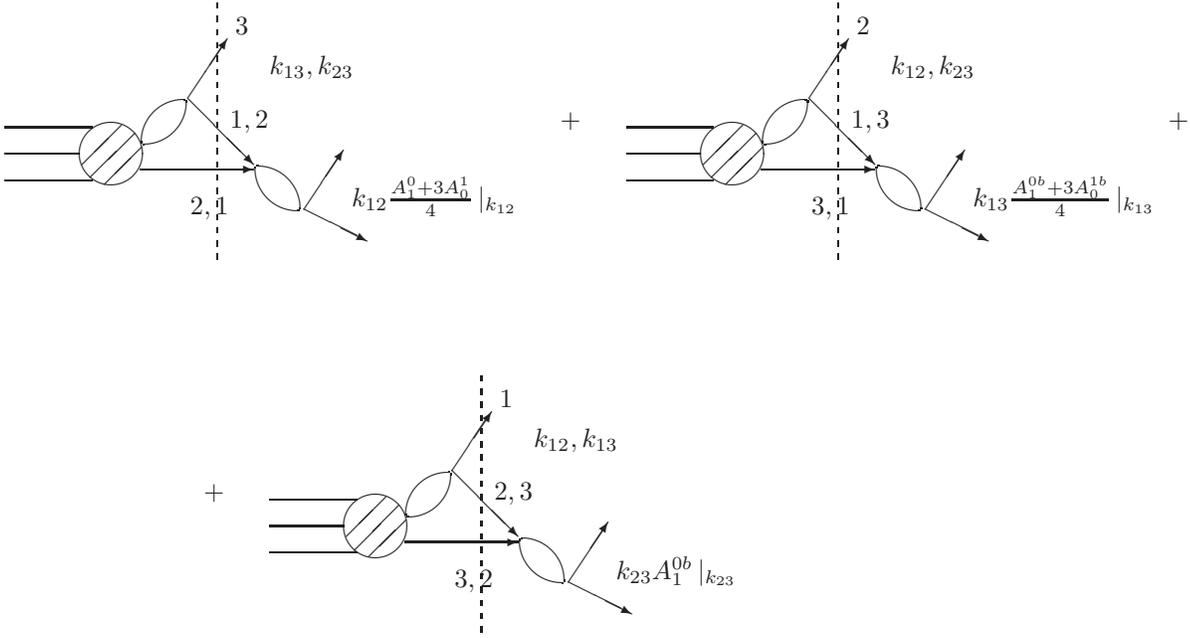
\put(-10,0){{\large Fig.3. The contribution of the diagrams with the
rescattering.}}
\end{picture}

%\newpage


\begin{thebibliography}{99}

\bibitem{1}
G.'t Hooft, Nucl. Phys. B{\bf 72}, 461 (1974).

\bibitem{2}
E. Witten, Nucl. Phys. B{\bf 160}, 57 (1979).

\bibitem{3}
I.J.R. Aitchison, J. Phys. G{\bf 3} 121 (1977).

\bibitem{4}
J.J. Brehm, Ann. Phys. (N.Y.) {\bf 108} 454 (1977).

\bibitem{5}
I.J.R. Aitchison and J.J. Brehm, Phys. Rev. D{\bf 17} 3072 (1978).

\bibitem{6}
I.J.R. Aitchison and J.J. Brehm, Phys. Rev. D{\bf 20} 1119, 1131 (1979).

\bibitem{7}
J.J. Brehm, Phys. Rev. D{\bf 21} 718 (1980).

\bibitem{8}
S.M. Gerasyuta, Yad. Fiz. {\bf 55} 3030 (1992).

\bibitem{9}
S.M. Gerasyuta, Z. Phys. C{\bf 60} 683 (1993).

\bibitem{10}
S.M. Gerasyuta and E.E. Matskevich, Yad. Fiz. {\bf 70} 1995 (2007).

\bibitem{11}
S.M. Gerasyuta and E.E. Matskevich, Int. J. Mod. Phys. E{\bf 17}, 585 (2008).

\bibitem{12}
G. Chew and S. Mandelstam, Phys. Rev. {\bf 119}, 467 (1960).

\bibitem{13}
A.De Rujula, H.Georgi and S.L.Glashow, Phys. Rev. D{\bf 12} 147 (1975).

\bibitem{14}
V.V. Anisovich, S.M. Gerasyuta and A.V. Sarantsev,
Int. J. Mod. Phys. A{\bf 6} 625 (1991).

\bibitem{15}
S.M. Gerasyuta and D.V. Ivanov, Nuovo Cim. A{\bf 112}, 261 (1999).


\end{thebibliography}
\end{document}